\documentclass[12pt]{article}

\usepackage[labelfont=bf,labelsep=period]{caption}

\usepackage{algorithmic}
\usepackage{amsmath, amssymb, dsfont,amsthm}
\usepackage{epsfig}
\usepackage{subfig}
\usepackage{xspace}
\usepackage{pstricks}

\newtheorem{thm}{Theorem}

\usepackage[numbers,sort&compress,longnamesfirst,sectionbib]{natbib}

\newcommand{\edge}{\rightarrow}
\newcommand{\path}{\leadsto}

\newcommand{\DAG}{{\ensuremath{\textit{DAG}}}\xspace}
\newcommand{\DAGn}{{\ensuremath{\textit{DAG}\,^n}}\xspace}

\newcommand{\di}{{\ensuremath{\delta^{(i)}}}\xspace}
\newcommand{\ddi}{{\ensuremath{|\di|}}\xspace}
\newcommand{\dddi}{{\ensuremath{\|\di\|}}\xspace}
\newcommand{\ddddi}{{\ensuremath{|\rangle \di \langle|}}\xspace}

\newcommand{\kkk}{{\ensuremath{\|K\|}}\xspace}
\newcommand{\kkkk}{{\ensuremath{|\rangle K\langle|}}\xspace}

\newcommand{\km}{{\ensuremath{\hat K}}\xspace}

\newcommand{\kkkkm}{{\ensuremath{|\rangle \km\langle|}}\xspace}

\newcommand{\Km}{{\ensuremath{\hat K^{(i)}}}\xspace}

\newcommand{\KKKKm}{{\ensuremath{|\rangle \Km\langle|}}\xspace}

\newcommand{\RF}{{\ensuremath{R_F^{(i)}}}\xspace}
\newcommand{\RB}{{\ensuremath{R_B^{(i)}}}\xspace}

\newcommand{\Ex}[1]{\ensuremath{\textup{\textbf{E}}\left[#1\right]}}
\newcommand{\Exx}[2]{\ensuremath{\textup{\textbf{E$_{#1}$}}\left[#2\right]}}
\newcommand{\Exxx}[3]{\ensuremath{\textup{\textbf{E$^{#1}_{#2}$}}\left[#3\right]}}
\newcommand{\ExM}[1]{\ensuremath{\textup{\textbf{E$_M$}}\left[#1\right]}}
\newcommand{\ExP}[1]{\ensuremath{\textup{\textbf{E$_p$}}\left[#1\right]}}
\newcommand{\bEx}[1]{\ensuremath{\textup{\textbf{E}}\left[#1\right]}}

\renewcommand{\O}{\ensuremath{{\cal O}}}

\newcommand{\inval}{\textsc{inval}}

\newcommand{\figref}[1]{Figure~\ref{fig:#1}}
\newcommand{\secref}[1]{Section~\ref{sec:#1}}

\newcommand{\secrefff}[2]{Sections~\ref{sec:#1}-\ref{sec:#2}}
\newcommand{\eq}[1]{Equation~\eqref{eq:#1}}

\newcommand{\thmref}[1]{Theorem~\ref{thm:#1}}
\newcommand{\thmrefs}[2]{Theorems~\ref{thm:#1} and~\ref{thm:#2}}

\newcommand{\ie}{i.\,e.\xspace}
\newcommand{\eg}{e.\,g.\xspace}

\DeclareSymbolFont{AMSb}{U}{msb}{m}{n}
\newcommand{\N}{{\mathbb{N}}}

\def\polylog{\operatorname{polylog}}

\allowdisplaybreaks[2]   

\title{\bf Average-Case Analysis of\\Online Topological Ordering\thanks{A conference version appeared in the 18th International Symposium on Algorithms and Computation (ISAAC 2007).}}

\author{Deepak Ajwani\footnotemark[2] \and Tobias Friedrich\thanks{Max-Planck-Institut f{\"u}r Informatik, Saarbr{\"u}cken, Germany}}
\date{}

\begin{document}

\maketitle

\begin{abstract}
\normalsize
\noindent
Many applications like pointer analysis and incremental compilation
require maintaining a topological ordering of
the nodes of a directed acyclic graph (DAG)
under dynamic updates.
All known algorithms for this problem are 
either only analyzed for worst-case insertion sequences or only evaluated 
experimentally on random DAGs.  We present the first average-case analysis of 
online topological ordering algorithms.  We prove an expected runtime of 
$\O(n^2\,\polylog(n))$ under insertion of the edges of a complete DAG in a random order 
for the algorithms of Alpern et al. (SODA, 1990), Katriel and Bodlaender (TALG, 2006),
and Pearce and Kelly (JEA, 2006).  This is much less than the best known 
worst-case bound $\O(n^{2.75})$ for this problem.
\end{abstract}

\section{Introduction}
There has been a growing interest in dynamic graph algorithms over the last two 
decades due to their applications in a variety of contexts including operating 
systems, information systems, network management, assembly planning, VLSI design 
and graphical applications. Typical dynamic graph algorithms maintain a certain 
property (\eg, connectivity information) of a graph that changes (a new edge 
inserted or an existing edge deleted) dynamically over time. An algorithm or a 
problem is called \textit{fully dynamic} if both edge insertions and deletions 
are allowed, and it is called \textit{partially dynamic} if only one (either 
only insertion or only deletion) is allowed. If only insertions are allowed, the 
partially dynamic algorithm is called incremental; if only deletions are 
allowed, it is called decremental. While a number of fully dynamic algorithms 
have been obtained for various properties on undirected graphs (see~\cite{DynGraphAlg}
and references therein), the design and 
analysis of fully dynamic algorithms for directed graphs has turned out to be 
much harder (\eg,~\cite{ZwickReachability,ZwickShortestPath,RR,Frigioni:SP}).
Much of the research on
directed graphs is therefore concentrated 
on the design of partially dynamic algorithms instead
(\eg,~\cite{Cicerone:semi-dynamic,Ausiello:MLP,Irit}).
In this paper, we focus 
on the analysis of algorithms for maintaining a topological ordering of directed 
graphs in an incremental setting.

A topological order $T$ of a directed graph $G = (V,E)$ (with $n:=|V|$ and 
$m:=|E|$) is a linear ordering of its nodes such that for all directed paths 
from $x \in V$ to $y \in V$ ($x \neq y$), it holds that $T(x) < T(y)$. A 
directed graph has a topological ordering if and only if it is acyclic. There 
are well-known algorithms for computing the topological ordering  of a directed 
acyclic graph (DAG) in $\O(m+n)$ time in an offline setting (see 
\eg~\citep{Cormen}). In a fully dynamic setting, each time an edge is added or 
deleted from the DAG, we are required to update the bijective mapping~$T$. In 
the online/incremental variant of this problem, the edges of the DAG are not known in advance but are inserted one at a time (no deletions allowed). 
As the topological order remains valid when removing edges, most algorithms
for online topological ordering can also handle the fully dynamic setting.
However, there are no good bounds known for the fully dynamic case.
Most algorithms are only analyzed in the online setting.

Given an arbitrary sequence of edges, the online cycle detection problem is to 
discover the first edge which introduces a cycle.  Till now, the best known 
algorithm for this problem involves maintaining an online topological order and 
returning the edge after which no valid topological order exists. Hence, results 
for online topological ordering also translate into results for the online cycle 
detection problem. Online topological ordering is required for incremental 
evaluation of computational circuits~\citep{Alpern} and in incremental 
compilation~\citep{MNR1,OLB} where a dependency graph between modules is 
maintained to reduce the amount of recompilation performed when an update 
occurs. An application for online cycle detection is pointer 
analysis~\citep{Cycle-Det}.

For inserting $m$ edges, the na\"{\i}ve way of computing an online topological order each time from scratch
with the offline algorithm takes $\O(m^2 + mn)$ time.
\citet{MNR} gave an algorithm that can insert $m$ edges in $\O(mn)$ time.
\citeauthor*{Alpern} (AHRSZ) proposed an algorithm~\citep{Alpern}
which runs in $\O(\kkkkm\,\log({\kkkkm}))$ time per edge insertion with
$\kkkkm$ being a local measure of the insertion complexity.
However, there is no analysis of AHRSZ for a sequence of edge insertions. 
\citeauthor*{Irit} (KB)~\citep{Irit} analyzed a variant of the AHRSZ algorithm 
and obtained an upper bound of $\O(\min\{m^{\frac{3}{2}} \log{n}, 
m^{\frac{3}{2}} + n^2 \log{n}\})$ for inserting an arbitrary sequence of $m$ 
edges. The algorithm by \citeauthor*{PK:JEA} (PK) \citep{PK:JEA} empirically outperforms 
the other algorithms for random edge insertions leading to sparse random DAGs, 
although its worst-case runtime is inferior to KB. \citeauthor*{SWAT06} (AFM)~\citep{SWAT06} proposed a new algorithm with runtime 
$\O(n^{2.75})$, which asymptotically outperforms KB on dense DAGs.

As noted above, the empirical performance on random edge insertion sequences 
(REIS) for the above algorithms are quite different from their worst-cases. 
While PK performs empirically better for REIS, KB and AFM are 
the best known algorithms for worst-case sequences. This leads us to the 
theoretical study of online topological ordering algorithms on REIS.
A nice property of such an average-case analysis is that (in contrast to worst-case bounds) the average of experimental results on REIS converge towards the real average after sufficiently many iterations.
This can give a good indication of the tightness of the proven theoretical bounds.

Our contributions are as follows:
\begin{itemize}
\item
    We show an expected runtime of $\O(n^2 \log^2{n})$ for inserting all
    edges of a complete DAG in a random order with PK (cf.~\secref{PK}).
\item
    For AHRSZ and KB, we show an expected runtime of $\O(n^2 \log^3{n})$ for complete random
    edge insertion sequences (cf.~\secref{AHRSZ}). This is significantly better 
    than the known worst-case bound of $\O(n^3)$ for KB to insert $\Omega(n^2)$ 
    edges. 
\item
    Additionally, we show that for such edge insertion sequences, the expected 
    number of edges which force any algorithm to change the topological order 
    (``invalidating edges'') is $\O(n^{\frac{3}{2}} \sqrt{\log{n}})$ 
    (cf.~\secref{inval}), which is the first such result.
\end{itemize}

The remainder of this paper is organized as follows.
The next section describes brief{}ly the three algorithms AHRSZ, KB, and PK.
In \secref{rand} we specify the random graph models used in our analysis.
\secrefff{PK}{inval} prove our upper bounds for the runtime of the three algorithms
and the number of invalidating edges.
\secref{exp} presents an empirical study,
which provides a deeper insight on the average case behavior of AHRSZ and PK.


\section{Algorithms}
\label{sec:algo}
This section first introduces some notations and then
describes the three algorithms AHRSZ, KB, and PK.
We keep the current topological order as a bijective function
$T\colon V \rightarrow [1..n]$.  In this and the subsequent sections, we will use the following notations:
$d(u,v)$ denotes $|T(u)-T(v)|$,
$u<v$ is a short form of $T(u)<T(v)$,
$u \edge v$ denotes an edge from~$u$ to~$v$, and
$u \path v$ expresses that $v$ is reachable from~$u$.
Note that $u \path u$, but \textit{not} $u \edge u$.  
The \emph{degree} of a node is the sum of its in- and out-degree.

Consider the $i$-th edge insertion $u\edge v$. We say that an edge insertion is 
\emph{invalidating} if $u>v$ before the insertion of this edge.
We define $\RB := \{x\in V \mid v\leq x \wedge x\path u\}$,
$\RF := \{y\in V \mid y \leq u \wedge v \path y\}$
and $\di = \RF \cup \RB$.
Let $\ddi$ denote the number of nodes in $\di$ and
let $\dddi$ denote the number of edges incident to nodes of $\di$. Note that 
$\di$ as defined above is different from the adaptive parameter~$\delta$ of the 
bounded incremental computation model. If an edge is non-invalidating, then 
$|\RB| = |\RF| = \ddi = 0$. Note that for an invalidating edge, $\RF \cap \RB = 
\emptyset$ as otherwise the algorithms will just report a cycle and terminate. 

We now describe the insertion of the $i$-th edge $u\edge v$ for all the three 
algorithms.  Assume for the remainder of this section that $u \edge v$ is an 
invalidating edge, as otherwise none of the algorithms do anything for that 
edge. We define an algorithm to be \emph{local} if it only changes the ordering 
of nodes~$x$ with $v\leq x\leq u$ to compute the new topological order~$T'$ of 
$G\cup\{(u,v)\}$. All three algorithms are local and they work in two phases -- a 
``discovery phase'' and a ``relabelling phase''.

In the discovery phase of \textbf{PK}, the set $\di$ is identified using a 
forward depth-first search from~$v$ (giving a set $\RF$) and a backward depth-
first search from~$u$ (giving a set $\RB$). The relabelling phase is also very 
simple.  It sorts both sets $\RF$ and $\RB$ separately in increasing topological 
order and then allocates new priorities according to the relative position in the 
sequence $\RB$ followed by $\RF$. It does not alter the priority of any node not 
in~$\di$, thereby greatly simplifying the relabeling phase. The runtime of PK 
for a single edge insertion is $\Theta(\dddi + \ddi \log{\ddi})$.

\citet{Alpern} used the bounded incremental computation model~\citep{RR}
and introduced the measure $\kkkkm$.
For an invalidated topological order~$T$, the set $K\subseteq V$  is a \emph{cover} if
for all $x,y\in V\colon (x\path y\ \wedge\ y<x\ \Rightarrow\ x\in K \vee y\in K)$.
This states that for any connected $x$ and $y$ which are incorrectly ordered, a cover
$K$ must include $x$ or $y$ or both.
$|K|$ and $\|K\|$ denote the number of nodes and edges touching nodes in~$K$, respectively.
We define $|\rangle K \langle| := |K| + \|K\|$ and a cover $\km$ to be 
\emph{minimal} if $\kkkkm \leq \kkkk$ for any other cover~$K$.
Thus, $\kkkkm$ captures the minimal amount of work required
to calculate the new topological order~$T'$ of $G\cup\{(u,v)\}$
assuming that the algorithm is local
and that the adjacent edges must be traversed.

\textbf{AHRSZ}s 
\emph{discovery phase} marks the nodes of a cover~$K$ by marking some of the
unmarked nodes $x,y\in\di$ with $x\path y$ and $y<x$.
This is done recursively by moving two frontiers starting from~$v$ and~$u$ towards each other.
Here, the crucial decision is which frontier to move next.
AHRSZ tries to minimize $\kkk$ by balancing the number of edges seen on both sides of the frontier.
The recursion stops when forward and backward frontier meet. Note that we do not 
necessarily visit all nodes in~$\RF$ ($\RB$) while extending the forward 
frontier (backward frontier). 
It can be proven \citep{Alpern} that the marked nodes indeed form a cover~$K$ and
that $\kkkk\leq 3\,\kkkkm$.

The \emph{relabeling phase} employs the dynamic priority space data structure
due to \citet{Dietz}.  This permits new priorities to be created between
existing ones in $\O(1)$ amortized time.  This is done in two passes over the nodes
in~$K$.  During the first pass, it visits the nodes of~$K$ in reverse topological order
and computes a strict upper bound on the new priorities
to be assigned to each node.
In the second phase, it visits the nodes in~$K$ in topological order and computes a strict
lower bound on the new priorities.
Both together allow to assign new priorities to each node in~$K$.
Thereafter they minimize the number of different labels used to speed up
the operations on the priority space data structure in practice.
It can be proven that the discovery phase with $\kkkkm$ priority queue operations dominates the time complexity,
giving an overall bound of $\O(\kkkkm \log{\kkkkm})$.

\textbf{KB} is a slight modification of AHRSZ.  In the discovery phase
AHRSZ counts the total number of edges incident on a node.  KB counts instead
only the in-degree of the backward frontier nodes and only the out-degree of
the forward frontier nodes. In addition, KB also simplified the relabeling phase.
The nodes visited during the extension of the forward (backward) frontier are deleted from the dynamic priority space data-structure and are reinserted, in the same relative order among themselves, after (before) all nodes in $\RB$ ($\RF$) not visited during the backward (forward) frontier extension.
The algorithm thus computes a cover $K \subseteq \di$ and its complexity per
edge insertion is $\O(\kkkk \log{\kkkk})$. The worst case running time of KB for a sequence of $m$ edge insertions is $\O(\min\{m^{\frac{3}{2}} \log{n}, m^{\frac{3}{2}} + n^2 \log{n}\})$.


\section{Random Graph Model}
\label{sec:rand}
\citet{Erdoes1,Erdoes2} introduced and popularized random graphs.
They defined two closely related models: $G(n,p)$ and $G(n,M)$.
The $G(n,p)$ model ($0<p<1$) consists of a graph with $n$ nodes in which each edge is chosen independently with probability $p$.
On the other hand, the $G(n,M)$ model
assigns equal probability
to all graphs with $n$ nodes and exactly $M$ edges. Each such graph occurs with a
probability of $1\big/\binom{N}{M}$, where $N:= \binom{n}{2}$. 

For our study of online topological ordering algorithms, we use the random DAG 
model of \citet{Barak}. They obtain a random DAG by directing the edges of an 
undirected random graph from lower to higher indexed vertices.
Depending on the underlying random graph model, this defines the $\DAG(n,p)$ and 
$\DAG(n,M)$ model. We will mainly work on the $\DAG(n,M)$  model since it is 
better suited to describe incremental addition of edges.

The set of all DAGs with $n$ nodes is denoted by $\DAGn$.
For a random variable~$f$ with probability space $\DAGn$,
$\ExM{f}$ and $\ExP{f}$ denotes
the expected value in the $\DAG(n,M)$ and $\DAG(n,p)$ model, respectively.
For the remainder of this paper, we set $\Ex{f} := \ExM{f}$ and $q:=1-p$.

The following theorem shows that in most investigations the models $\DAG(n,p)$ and $\DAG(n,M)$
are practically interchangeable, provided $M$ is close to $p N$.
\begin{thm}
    \label{thm:rand_mod_dag}
    Given a function $f\colon \DAGn \rightarrow [0,a]$ with $a>0$ and
    $f(G) \leq f(H)$ for all $G \subseteq H$
    and functions $p$ and $M$ of $n$ with $0 < p < 1$ and $M \in \N$.
    \begin{enumerate}
        \setlength{\itemsep}{0pt}
        \setlength{\parskip}{0pt}
        \item
        If\
        $\displaystyle
            \lim_{n \rightarrow \infty} p q N =
            \lim_{n \rightarrow \infty} \frac{p N - M}{\sqrt{p q N}} 
            = \infty,
        $
        \ then\ \
        $\displaystyle
            \ExM{f} \leq \ExP{f} + o(1).
        $
        \item
        If\ 
        $\displaystyle
            \lim_{n \rightarrow \infty} p q N =
            \lim_{n \rightarrow \infty} \frac{M - p N}{\sqrt{p q N}} 
            = \infty,
        $
        \ then\ \
        $\displaystyle
            \ExP{f} \leq \ExM{f} + o(1).
        $
    \end{enumerate}
\end{thm}
The analogous theorem for the undirected graph models $G(n,p)$ and $G(n,M)$ is well known.
A closer look at the proof for it given by \citet{BollobasRand} reveals that the
probabilistic argument used to show the close connection between
$G(n,p)$ and $G(n,M)$ can be applied in the same manner for the two random DAG models
$\DAG(n,p)$ and $\DAG(n,M)$.

We define a random edge sequence to be a uniform random permutation of the edges 
of a complete DAG, \ie, all permutations of $\binom{n}{2}$ edges are equally 
likely. If the edges appear to the online algorithm in the order in which they 
appear in the random edge sequence, we call it a random edge insertion sequence 
(REIS). Note that a DAG obtained after inserting $M$ edges of a REIS will have 
the same probability distribution as $\DAG(n,M)$. To simplify the proofs, we 
first show our results in $\DAG(n,p)$ model and then transfer them in the 
$\DAG(n,M)$ model by \thmref{rand_mod_dag}.


\section{Analysis of PK}
\label{sec:PK}

When inserting the $i$-th edge $u\edge v$, PK only regards nodes in
$\di:=\{x\in V \mid v\leq x\leq u \wedge (v\path x \vee x\path u)\}$
with ``$\leq$'' defined according to the current topological order. As discussed 
in \secref{algo}, PK performs $\O(\dddi + \ddi \log{\ddi})$ operations for 
inserting the $i$-th edge. The intuition behind the proofs in this section is 
that in the early phase of edge-insertions (the first $O(n \log{n})$ edges), the 
graph is sparse and so only a few edges are traversed during the DFS traversals. 
As the graph grows, fewer and fewer nodes are visited in DFS traversals ($\ddi$ 
is small) and so the total number of edges traversed in DFS traversals (bounded 
above by $\dddi$) is still small.

\thmrefs{delta1}{delta3} of this section show for
a random edge insertion sequence (REIS) of $N$~edges that
$\sum_{i=1}^{N}{\ddi} = \O(n^2)$ and
$\bEx{\sum_{i=1}^{N}{\dddi}} = \O (n^2 \log^2{n})$.
This proves the following theorem. 
\begin{thm}
\label{thm:PK}
For a random edge insertion sequence (REIS) leading to a complete DAG,
the expected runtime of PK is $\O(n^2 \log^2{n})$.
\end{thm}

A comparable pair (of nodes) are two distinct nodes $x$ and $y$ such that either 
$x \leadsto y$ or $y \leadsto x$.  We define a potential function $\Phi_i$ 
similar to Katriel and Bodlaender~\cite{Irit}. Let $\Phi_i$ be the number of 
comparable pairs after the insertion of $i$ edges. Clearly,
\begin{equation}
\begin{minipage}[c]{220pt}
    $\Delta\Phi_i:=\Phi_i-\Phi_{i-1}\geq 0$ \quad for all $1\leq i \leq M$,\\
    \hspace*{.7cm}$\Phi_0=0$,\quad and\quad  $\Phi_M \leq n(n-1) / 2$.
\end{minipage}\label{eq:phi1}
\end{equation}

\begin{thm}
\label{thm:di}
For all edge sequences, (i) $\displaystyle \ddi \leq \Delta\Phi_i+1$ and (ii) $\displaystyle \ddi \leq 2\Delta\Phi_i$.
\end{thm}
\begin{proof}
  Consider the $i$-th edge $(u,v)$. 
    If $u < v$, the theorem is trivial since $\ddi=0$.
    Otherwise, each vertex of \RF\ and \RB\
    (as defined in \secref{algo})
    gets newly
    ordered with respect to $u$ and $v$, respectively. The set 
    $\bigcup_{x\in\RB}(x,v) \cap \bigcup_{x\in\RF}(u,x)=\{(u,v)\}$.
    This means that overall at least $|\RF|+|\RB|-1$ node pairs get newly ordered:
    \[
      \Delta\Phi_i \geq |\RF|+|\RB| -1 = \ddi -1.
    \]
    Also, since in this case $\Delta\Phi_i \geq 1$, $\ddi \leq 2\Delta\Phi_i$.
\end{proof}

\begin{thm}
    \label{thm:delta1}
    For all edge sequences,  $\displaystyle\sum_{i=1}^{N}{\ddi} \leq n(n-1) = \O(n^2)$.
\end{thm}
\begin{proof}
    By \thmref{di} (i), we get
    $\displaystyle
        \sum_{i=1}^{N}{\ddi}
          \leq \sum_{i=1}^{N}(\Delta \Phi_i+1) = \Phi_N + N \leq
          n(n-1)/2 + n(n-1)/2 = n(n-1).$
\end{proof}

The remainder of this section provides the necessary tools step by step to finally prove the
desired bound on $\sum_{i=1}^{N}{\dddi}$ in \thmref{delta3}.
One can also interpret $\Phi_i$ as a random variable in $\DAG(n,M)$ with $M=i$.
The corresponding function $\Psi$ for $\DAG(n,p)$
is defined as the total number of comparable node pairs in $\DAG(n,p)$.
\citet{RandomDAG} showed the following theorem.
\begin{thm}
\label{thm:pittel}
    For $p:=c \log(n)/n$ and $c > 1 $, $\ExP{\Psi}  = (1+o(1))\, \frac{n^2}{2} \left(1-\frac{1}{c}\right)^2$.
\end{thm}

Using \thmref{rand_mod_dag}, this result can be transformed to $\Phi$ as defined above for $\DAG(n,M)$
and gives the following bounds for $\ExM{\Phi_{k}}$.
\begin{thm}
\label{thm:phi_bound}
    For $n \log{n} < k \leq N - 2 n \log{n}$,
    \[
        \ExM{\Phi_{k}} = (1+o(1))\, \frac{n^2}{2} \left(1-\frac{(n-1)\log{n}}{2(k + n \log{n})}\right)^2.
    \]
    For $N - 2 n \log{n} < k \leq N - 2 \log{n}$,
    \[
        \ExM{\Phi_{k}} = (1+o(1)) \frac{n^2}{2} \!\left(1-\frac{(n -1)\log{n}}{2(k + \sqrt{\log{n}\,(N-k)})}\right)^2.
    \]
\end{thm}
\begin{proof}
    The function $\Psi\colon DAG^n \rightarrow [0,N]$ and $\Psi(G) \leq \Psi(H)$ 
    wherever $G \subseteq H$. The later inequality is true as the nodes already 
    ordered in $G$ will still remain ordered in~$H$. 
    For $n \log{n} < k \leq N - 2 n \log {n}$, consider $p := \frac{k+n\log{n}}{N}$.  Then
    \[
        \lim_{n \rightarrow \infty} p q N
        \geq \lim_{n \rightarrow \infty} \frac{\log{n}}{n} \frac{\log{n}}{n} N
        \geq \lim_{n \rightarrow \infty} \frac{(n-1) \log^2{n}}{2 n}
        = \infty
    \]
    and
    \begin{align*}
        \lim_{n \rightarrow \infty} \frac{pN - k}{\sqrt{pqN}}
        &\geq \lim_{n \rightarrow \infty} \frac{pN - k}{\sqrt{N}}
        \geq \lim_{n \rightarrow \infty} \frac{n \log{n}}{\sqrt{N}}\\
        &\geq \lim_{n \rightarrow \infty} \frac{n \log{n}}{n}
        \geq \lim_{n \rightarrow \infty} \log{n}
        = \infty.
    \end{align*}
    Since all the conditions of \thmref{rand_mod_dag} are satisfied
    for these values of $k$ and $p$, $\ExM{\Psi} = O(\ExP{\Psi})$. In particular,
    \[
        \ExM{\Phi_{k}}
         = \Exx{p=(k+n \log{n})/N}{\Psi} + o(1)
        = (1+o(1)) \frac{n^2}{2} \left(1-\frac{(n -1)\log{n}}{2(k + n \log{n})}\right)^2.
    \]
    For 
    $N - 2 n \log{n} < k \leq N - 2 \log{n}$,
    we choose 
    $p := \frac{k+\sqrt{\log{n}\,(N-k)}}{N}$. Clearly,
\[
    p \geq \frac{N - 2 n \log{n} + \sqrt{\log{n} (N - (N - 2 \log{n}))}}{N}
    \geq \frac{N - 2 n \log{n} + \sqrt{2} \log{n}}{N}.
\]
 Using this, we get
    \[
        \lim_{n \rightarrow \infty} p q N
        \geq \lim_{n \rightarrow \infty} \frac{(N-2n\log{n} +\sqrt{2}\log{n})}{N}\,
                                         \frac{(N-k-\sqrt{\log{n}\,(N-k)})\,N}{N}.
    \]
    Observe that $f(k):=N-k-\sqrt{\log{n}\,(N-k)}$ has its minimum
    at $k_0=N-\log(n)/4$ since
    $f'(k_0)=0$ and
    $f''(k_0)=2/\log{n}>0$.
    Hence, we conclude that $f(k)$ is monotonically decreasing in our interval $(N - 2 n \log{n},N - 2 \log{n})$
    and attains its minimum at $N - 2 \log{n}$.
    Therefore, $N-k-\sqrt{\log{n}\,(N-k)} \geq 2 \log{n} - \sqrt{2}\log{n}\rightarrow\infty$,
    which in turn proves $\lim_{n \rightarrow \infty} p q N = \infty$ and
    \[
        \lim_{n \rightarrow \infty} \frac{pN - k}{\sqrt{pqN}}
        \geq \lim_{n \rightarrow \infty} \frac{\sqrt{\log{n}\,(N-k)}}{\sqrt{N-k-\sqrt{\log{n}\,(N-k)}}}
        \geq \lim_{n \rightarrow \infty} \sqrt{\log{n}}
        = \infty
    \]
    Together with \thmref{pittel}, this yields
    \begin{align*}
        \ExM{\Phi_{k}}
        &= \Exx{p=(k+\sqrt{\log{n}\,(N-k)})/N}{\Psi} + o(1)\\
        &= (1+o(1))\, \frac{n^2}{2} \left(1-\frac{(n-1)\log{n}}{2(k + \sqrt{\log{n}\,(N-k)})}\right)^2.
        \qedhere
    \end{align*}
\end{proof}

The degree sequence of a random graph is a well-studied problem.
The following theorem is shown in \cite{BollobasRand}.
\begin{thm}
    If $pn/\log{n} \rightarrow \infty$, then almost every graph $G$ in the 
    $G(n,p)$ model satisfies $\Delta(G) = (1+o(1))\,pn$, where $\Delta(G)$ is 
    the maximum degree of a node in~$G$.
\end{thm} 

As noted in \secref{rand}, the undirected graph obtained by ignoring the 
directions of $\DAG(n,p)$ is a $G(n,p)$ graph. Therefore, the above result is 
also true for the maximum degree (in-degree + out-degree) of a node in 
$\DAG(n,p)$. Using \thmref{rand_mod_dag}, the above result can be transformed to 
$\DAG(n,M)$, as well. 

\begin{thm}
\label{thm:degree1}
    With probability $1-\O(\frac{1}{n})$, there is no node with degree 
    higher than $21 \frac{M}{n}$ for sufficiently large $n$ and $M > n \log{n}$ in 
    $\DAG(n,M)$.
\end{thm}
\begin{proof}
    We examine the following two functions:
    \begin{itemize}
    \setlength{\itemsep}{0pt}
    \setlength{\parskip}{0pt}
    \item $f_1(g)\colon$ Number of nodes with degree at least $g(n)$
    \item $f_2(g) := f_1^2(g)$
    \end{itemize}
    
    For $f_1,f_2$ in $G(n,p)$,
    $g(n):=p n + 2 \sqrt{p q n \log{n}}$, and some constant $c$,
    \citet{BollobasDegSeq} showed
    \begin{equation}
    \begin{minipage}[c]{200pt}
    \begin{align}
        \ExP{f_1(g)} &= \textstyle\O\left(\frac{1}{n}\right),\notag\\
        \sigma^2_p(f_1(g)) & = \ExP{f_2(g)} - \Exxx{2}{p}{f_1(g)} \leq c \cdot \ExP{f_1(g)}.\notag
    \end{align}
    \end{minipage}\label{eq:bollo_deg}
    \end{equation}
    
    Consider any random $\DAG(n,M)$. It must have been obtained by taking a random 
    graph $G(n,M)$ and ordering the edges. The degree of a node in $\DAG(n,M)$ is 
    the same as the degree of the corresponding node in $G(n,M)$.
    
    We break down the analysis depending on $M$.  At first, consider the simpler 
    case of $M > \left(\lfloor\frac{N}{n \log{n}}\rfloor - 2\right) n \log{n}$. The 
    degree of any node in an undirected graph cannot be higher than $n-1$. However, 
    as $M > N - 3 n \log{n}$, $21 \cdot \frac{M}{n} \geq \frac{21}{2} (n-1) - 63 
    \log{n}$. For sufficiently large $n$ this is greater than $n-1$ and therefore, 
    no node can have degree higher than it.
    
    Next, we consider $M \in (k n \log{n},(k+1)\, n \log{n}]$ for $1 \leq k < l$,
    where $l := \lfloor\frac{N}{n \log{n}}\rfloor - 2$, and we prove the theorem for each interval. We choose
    $p_k := (k+2) \frac{n \log{n}}{N}$,
    $q_k : = 1 - p_k$, and
    $g_k(n) := p_k n + 2 \sqrt{p_k q_k n \log{n}}$
    and look for the conditions in \thmref{rand_mod_dag}.
    Note that $0<p_k<1$, $f_1\colon G^n \rightarrow [0,n]$, $f_2\colon G^n \rightarrow 
    [0,n^2]$,  and $f_i(G) \leq f_i(H)$ wherever $G \subseteq H$ for $i=1,2$. The 
    later inequality holds as the degree of any node in $H$ is greater than or equal 
    to the corresponding degree in $G$.
    For $1 \leq k < l$,
    \[
        p_k \geq \frac{3 n \log{n}}{N}
        \geq\frac{6 \log{n}}{n-1}
    \]
    and
    \[
        q_k 
        \geq 1 - \left(\left\lfloor\frac{N}{n \log{n}}\right\rfloor-1\right) \frac{n \log{n}}{N}
        \geq 1 - \left(\frac{N- n \log{n}}{n \log{n}}\right) \frac{n \log{n}}{N}
        \geq \frac{2 \log{n}}{n-1}.
    \]
    
    So for each interval, 
    \[
        \lim_{n \rightarrow \infty} p_k q_k N
        \geq \lim_{n \rightarrow \infty} \frac{6\log{n}}{n-1} \frac{2\log{n}}{n-1} N
            \geq \lim_{n \rightarrow \infty} 6 \textstyle \log^2{n}
        = \infty
    \]
    and by $M_k \leq (k+1)\, n \log{n}$ and $\textstyle k+2 \leq \lfloor\frac{N}{n \log{n}}\rfloor$,
    \begin{align*}
    \lim_{n \rightarrow \infty} \frac{pN - M}{\sqrt{pqN}}
        &\geq \lim_{n \rightarrow \infty} \frac{pN - M}{\sqrt{pN}}
        \geq \lim_{n \rightarrow \infty} \frac{n \log{n}}{\sqrt{(k+2)\,n \log{n}}}
        = \lim_{n \rightarrow \infty} \frac{\sqrt{n \log{n}}}{\sqrt{k+2}}\\
       &\geq \lim_{n \rightarrow \infty} \frac{n \log{n}}{\sqrt{N}}
        \geq \lim_{n \rightarrow \infty} \log{n}
        = \infty
    \end{align*}
    In each interval, all the conditions of \thmref{rand_mod_dag}
    are satisfied and therefore, $\ExM{f_i(g_k)} = \Exx{p_k}{f_i(g_k)}+o(1)$
    for $i=1,2$ and $1 \leq k < l$. Using \eq{bollo_deg}, we get 
    $\ExM{f_1(g_k)} = \O(\Exx{p_k}{f_1(g_k)}) = \O\left(\frac{1}{n}\right)$ and 
    \begin{align*}
    \sigma_M^2(f_1(g_k)) &= \ExM{f_2(g_k)} - \Exxx{2}{M}{f_1(g_k)}
        = \O\big(\Exx{p_k}{f_2(g_k)} - \Exxx{2}{p_k}{f_1(g_k)}\big)\\
        &= \O(\sigma_{p_k}^2(f_1(g_k)))
        = \O(\Exx{p_k}{f_1(g_k)})
        = \O\left(\tfrac{1}{n}\right).
    \end{align*}
    Therefore, by substituting
    $X:=f_1(g_k)$,
    $\mu := \ExM{f_1(g_k)} = \O\left(\frac{1}{n}\right)$,
    $\sigma^2 := \sigma_M^2(f_1(g_k)) = \O\left(\frac{1}{n}\right)$, and
    $\nu := 1 - \mu$
    in Chebyshev's inequality ($\Pr\{|X-\mu| \geq \nu\} \leq \frac{\sigma^2}{\nu^2}$), we get 
    \begin{align*}
    \Pr\{|f_1(g_k) - \mu| \geq 1 - \mu\}& \leq \O\left(\frac{1}{n (1-\mu)^2}\right)
     = \O\left(\tfrac{1}{n}\right).
    \end{align*}
    However, $\Pr\{|f_1(g_k) - \mu| \geq 1 - \mu\} = \Pr\{(f_1(g_k) \geq 1) \text{ or\ }
    (f_1(g_k) \leq 2 \mu - 1)\}$ and since, $\mu = \O\left(\frac{1}{n}\right)$ and 
    $f_1(g_k)$ is non-negative random variable, $\Pr\{f_1(g_k) \leq 2 \mu - 1\} = 0$ 
    for sufficiently large $n$. Therefore, $\Pr\{f_1(g_k) \geq 1\} = \Pr\{|f_1(g_k) 
    - \mu| \geq 1 - \mu\} = \O\left(\textstyle\frac{1}{n}\right)$. In other words, 
    with probability $(1-\O(\frac{1}{n}))$, there is no node with a degree 
    higher than $g_k$ in any interval. However, by $p_k \geq 
    \textstyle\frac{\log{n}}{n}$ we get
    \[
    g_k(n) = p_k n + 2 \sqrt{p_k q_k n \log{n}}
        \leq 3 p_k n
        \leq 6 (k+2) \frac{n\ \log{n}}{n - 1}
    \]
    For sufficiently large $n$, $\frac{n}{n-1} \leq \frac{7}{6}$, and this implies
    \[    
    g_k(n) \leq 7 (k+2) \log{n}
        \leq \frac{7(k+2)}{k} \frac{M}{n}
        \leq \frac{21 M}{n}.
    \]
    Therefore, with probability $1-\O(\frac{1}{n})$, there is no node with 
    a degree higher than $21 \frac{M}{n}$ in $G(n,M)$ and by the argument above, in 
    $\DAG(n,M)$.
\end{proof}

As the maximum degree of a node in $\DAG(n,i)$ is $\O(i/n)$,
we finally just need to show a bound on $\sum_i{(i \cdot \ddi)}$
to prove \thmref{delta3}.
This is done in the following theorem.

\begin{thm}
    \label{thm:delta2}
    For $\DAG(n,M)$ and $r := N - 2 \log{n}$, 
    \[
        \displaystyle \bEx{\sum_{i=1}^{r}{(i \cdot \ddi)}}
                   = \O (n^3 \log^2{n}).
    \]
\end{thm}
\begin{proof}
    Let us decompose the analysis in three steps.
    First, we show a bound on the first $n\log{n}$
    edges. By definition of $\di$, $\ddi\leq n$. Therefore, 
    \begin{align}
        \sum_{i=1}^{n \log{n}}{i \cdot \Ex{\ddi}}
        \leq \sum_{i=1}^{n \log{n}}{i \cdot n}
        &=\O \left(n^3 \log^2{n}\right).\label{eq:sie1}
    \end{align}
    The second step is to bound $\sum_{i=n\log{n}}^{t}{i \cdot \ddi}$ with
    $t := N - 2 n \log{n}$.
    For this, \thmref{di}~(ii) shows
    for all $k$ such that $n \log{n} < k < t$ that
    \begin{align}
        \bEx{\sum_{i=k}^{t}{\ddi}}
        &\leq 2\,\Ex{\sum_{i=k}^{t}{\Delta \Phi_i }}
        = 2\,\Ex{\Phi_t - \Phi_{k-1}}
        = 2\,\Ex{\Phi_t} - 2\Ex{\Phi_{k-1}}.
        \label{eq:eddi}
    \end{align}
    The function hidden in the $o(1)$ in \thmref{pittel}
    is decreasing in $p$ \cite{RandomDAG}.
    Hence, also the $o(1)$ in \thmref{phi_bound} is
    decreasing in $k$.
    Plugging this in \eq{eddi}
    yields (with $s: = n\log{n}$)
    \begin{align}
        \bEx{\sum_{i=k}^{t}{\ddi}}
        &\leq (1+o(1))\, n^2 
               \Bigg(\Big(1-\frac{(n-1)\log{n}}{2(t+ s)}\Big)^2 -
                     \Big(1-\frac{(n-1)\log{n}}{2(k-1+s)}\Big)^2\Bigg)\notag\\
        &= (1+o(1))\, n^2 (n-1) \log{n} \Big(\frac{2}{2(k-1+s)} - \frac{2}{2(t + s)} \,+\notag\\
        &\hspace*{3.7cm}
               \frac{(n-1)\log{n}}{4}\Big(\frac{1}{(t+s)^2}-\frac{1}{(k-1+s)^2}\Big) \Big)\notag\\
        &\leq (1+o(1))\, n^2 (n-1) \log{n}
               \left(\frac{1}{k-1+s} - \frac{1}{t+s}\right)\notag\\
        &\leq
        (1+o(1))\, n^2 (n-1) \log{n} \frac{1}{k-1}.\label{eq:epk2}
    \end{align}

    By linearity of expectation and \eq{epk2},
    \begin{align*}
        \bEx{\sum_{i=s+1}^t i\,\ddi}
            & = \sum_{i=s+1}^t \Big( i\, \Ex{\ddi} \Big)
            \leq \sum_{j=1}^{\log{(\lceil \frac{t}{s} \rceil)}}
                \Big(2^j s \sum_{i=2^{(j-1)} s +1}^{2^j s} \Ex{\ddi} \Big)\\
            & \leq \sum_{j=1}^{\log{(\lceil \frac{t}{s} \rceil)}}
                \Big(2^j s \!\sum_{i=2^{(j-1)}s+1}^{t} \Ex{\ddi} \Big)\\
            & \leq \sum_{j=1}^{\log{(\lceil \frac{t}{s} \rceil)}}
                \Big(2^j s (1+o(1))\, n^2 (n-1) \log{n} \frac{1}{2^{(j-1)}s}\Big)\\
            & = \sum_{j=1}^{\log{(\lceil \frac{t}{s} \rceil)}}
                \big(2 (1+o(1))\, n^2 (n-1) \log{n}\big) \\
            & = 2 (1+o(1))\, n^2 (n-1) \log^2{n} = \O(n^3 \log^2{n}).
          \end{align*}
    \enlargethispage{-\baselineskip}
    For the last step consider a $k$ such that $t < k < r$.
    \thmref{di}~(ii) gives
    \begin{align*}
        \bEx{\sum_{i=k}^{r}{\ddi}}
        &\leq 2\,\bEx{\sum_{i=k}^{r}{\Delta \Phi_i }}
        = 2\,\Ex{\Phi_r - \Phi_{k-1}}
        = 2\,\Ex{\Phi_r} - 2\Ex{\Phi_{k-1}}.
    \end{align*}
    Using \thmref{phi_bound} and similar arguments as before, this yields (with $s(k): = \sqrt{\log{n}\ (N-k)}$)
    \begin{align*}
        \bEx{\sum_{i=k}^{r}{\ddi}}\hspace*{-1.1cm} \\
        &\leq (1+o(1))\,n^2 
               \Bigg(\Big(1-\frac{(n - 1)\log{n}}{2(r+ s(r))}\Big)^2 -
                     \Big(1-\frac{(n - 1)\log{n}}{2(k-1+s(k-1))}\Big)^2\Bigg)\\
        &= (1+o(1))\,n^2 (n-1) \log{n} \Bigg(\frac{2}{2(k-1+s(k-1))} - \frac{2}{2(r + s(r))} \,+ \\
        &\hspace*{3.2cm}
               \frac{(n - 1)\log{n}}{4}\Big(\frac{1}{(r+s(r))^2}-\frac{1}{(k-1+s(k-1))^2}\Big) \Bigg).
    \end{align*}
    Since $k+s(k)$ is monotonically increasing for $t<k<r$, $\tfrac{1}{(k+s(k))^{2}}$ is a monotonically decreasing function in this interval.  Therefore, $\frac{1}{(r+s(r))^2}-\frac{1}{(k-1+s(k-1))^2} < 0$, which proves the following equation.
    \begin{align}
        \bEx{ \sum_{i=k}^{r}{\ddi} } &\leq (1+o(1))\, n^2 (n-1) \log{n}
               \left(\frac{1}{k-1+s(k-1)} - \frac{1}{r+s(r)}\right)\notag\\
        &\leq
        (1+o(1))\,n^2 (n-1) \log{n} \frac{1}{k-1}.\label{eq:epk52}
    \end{align}
    By linearity of expectation and \eq{epk52},
        \begin{align*}
            \bEx { \sum_{i= N - 2 n \log{n} +1}^r i\,\ddi }\hspace*{-3.1cm} \\
                & = \sum_{i= N - 2 n \log{n} +1}^r \Big( i\, \Ex{\ddi} \Big)\notag \\
                & \leq (N - 2 \log{n})\, \sum_{i=N-2 n \log{n}+1}^r \Ex{\ddi}\notag\\
                & \leq (N - 2 \log{n})\, (1+o(1))\, n^2 (n-1) \log{n} \frac{1}{N - 2 n \log{n} -1}\notag\\
                & = \O(n^3 \log{n}). \notag
            \qedhere
          \end{align*}
\end{proof}

\begin{thm}
    \label{thm:delta3}
    For $\DAG(n,M)$, $\displaystyle\bEx{\sum_{i=1}^{N}{\dddi}}= \O (n^2 \log^2{n})$.
\end{thm}
\begin{proof}
    By definition of \dddi, we know $\dddi \leq i$ and hence
    \[\sum_{i=1}^{n \log{n}} \dddi = \O(n^2 \log^2{n}).\]
    
    Again, let $r := N - 2 \log{n}$. \thmref{degree1}
    tells us that with probability greater than $\big(1-\frac{c'}{n}\big)$
    for some constant $c'$, there is no node with degree $\geq \frac{c\ i}{n}$ (for $c = 21$).
    Since the degree of an arbitrary node in a DAG is bounded by $n$, 
    we get with \thmrefs{delta1}{delta2},
    \begin{align*}
        \bEx{\sum_{i=n \log{n} +1}^{r}{\dddi}}
        &= \O\Bigg(\bEx{\sum_{i=n \log{n} + 1}^{r}\! {\frac{c\ i\ \ddi}{n} }} +
                   \bEx{\sum_{i=n \log{n} + 1}^{r}\! \frac{n\ c'\ \ddi}{n} }\Bigg)\\
        &= \O \Big(\frac{1}{n}\, \bEx{\sum_{i=1}^{r}{(i \ \ddi)}} + n^2\Big)\\
        &= \O\Big( \frac{1}{n} \left(n^3 \log^2{n}\right) + n^2 \Big)
        = \O (n^2 \log^2{n}).
    \end{align*}
By again using the fact that the degree of an arbitrary node in a DAG is at most~$n$, we obtain
\[
    \bEx{\sum_{i=r +1}^{N}{\dddi}}
    = \O\Big(n \cdot \bEx{\sum_{i=r +1}^{N}{\ddi}}\Big)
    = \O\Big(n \cdot \sum_{i=r +1}^{N} n \Big)
    = \O(n^2 \log{n}).
\]
Thus,
\begin{align*}
    \bEx{\sum_{i=1}^{N}{\dddi}}
    & = \bEx{\sum_{i=1}^{n\log{n}}{\dddi}} +
        \bEx{\sum_{i=n\ \log{n}+1}^{r}{\dddi}} +
        \bEx{\sum_{i=r+1}^{N}{\dddi}}\\
    & = \O(n^2 \log^2{n}) + \O(n^2 \log^2{n}) + \O(n^2 \log{n}) = \O(n^2\ \log^2{n}).
    \qedhere
\end{align*}
\end{proof}


\section{Analysis of AHRSZ and KB}
\label{sec:AHRSZ}
\citet{Irit} introduced KB as a variant of AHRSZ for which a worst-case runtime 
of $\O(\min\{m^{\frac{3}{2}} \log{n}, m^{\frac{3}{2}} + n^2 \log{n}\})$ can be 
shown. In this section, we prove an expected runtime of $\O(n^2 \log^3{n})$ 
under random edge insertion sequences,
both for AHRSZ and KB.

Recall from \secref{algo} that
for every edge insertion there is a minimal cover \Km.
The following theorem shows that $\di$ is also a valid cover in this situation.

\begin{thm}
\label{thm:del_kmin}
    $\di$ is a valid cover.
\end{thm}
\begin{proof}
    Consider the insertion of the $i$-th edge $(u,v)$ and consider a node-pair $x,y$ 
    such that $x \path y$, but $x > y$. Since before the insertion of this edge, the 
    topological ordering was consistent, $x \path u \edge v \path y$, $x < u$ and
    $v < y$. Together with $x > y$, it implies $x > v$. Now $x \path u$ and $x \geq v$ 
    imply $x \in \di$. Thus, for every node-pair $(x,y)$ such that $x \path y$ and
    $x > y$, $x \in \di$ and hence, $\di$ is a valid cover.
\end{proof}

Therefore, by definition of $\KKKKm$, $\KKKKm \leq \ddddi = \ddi + \dddi$.
\begin{equation*}
    \bEx{\sum_{i=1}^m \KKKKm}
        \leq \sum_{i=1}^m  \ddi + \bEx{\sum_{i=1}^m \dddi}
        = \O(n^2 \log^2{n})\\
\end{equation*} 
The latter equality follows from \thmrefs{delta1}{delta3}. The expected
complexity of AHRSZ on REIS is thus $\O\big(\bEx{\sum_{i=1}^m \KKKKm \log{n}}\big) = \O(n^2 \log^3{n})$.

\begin{sloppypar}
KB also computes a cover $K \subseteq \di$ and its complexity per edge insertion is
$\O(\kkkk\, \log{\kkkk})$. Therefore, $\kkkk \leq \ddi + \dddi$ and with a similar
argument as above, the expected complexity of KB on REIS is $\O(n^2 \log^3{n})$.
\end{sloppypar}


\section{Bounding the number of invalidating edges}
\label{sec:inval}
An interesting question in all this analysis is how many edges will actually 
invalidate the topological ordering and force any algorithm to do 
something about them. Here, we show a non-trivial upper bound on the expected 
value of the number of invalidating edges on REIS. Consider the following random 
variable: $\inval(i)=1$ if the $i$-th edge inserted is an invalidating edge; $\inval(i)=0$ otherwise.

\begin{thm}
    \label{thm:inval}
    $\displaystyle\bEx{ \sum_{i=1}^m {\inval(i)} } = \O(\min\{m,n^{\frac{3}{2}} \log^{\frac{1}{2}}{n}\}).$
\end{thm}
\begin{proof}
 If the $i$-th edge is invalidating, $\ddi \geq 2$; otherwise $\inval(i)=\ddi=0$. In either case, $\inval(i) \leq \ddi/2$. Thus, for $s:=n^\frac{3}{2} \log^\frac{1}{2}{n}$ and $t := \min\{m,N - 2 n \log{n}\}$,
    \begin{align*}
        \bEx{ \sum_{i=s +1}^{t}{\inval(i)}}
            &\leq \bEx { \sum_{i=s +1}^{t}{\frac{\ddi}{2}} }
             \leq (1+o(1))\,\frac{n^2 (n-1) \log{n}}{2 s}\\
            &\leq \frac{(1+o(1))}{2}\,n^\frac{3}{2} \log^\frac{1}{2}{n}.
    \end{align*}
The second inequality follows by substituting $k:=s +1$ in \eq{epk2}.
    Also, since the number of invalidating edges can be at most equal to the total 
    number of edges, \mbox{$\sum_{i=1}^{s} \inval(i) \leq s$}.
    \begin{align*}
        \bEx { \sum_{i=1}^m {\inval(i)} }
            &= \bEx { \sum_{i=1}^{s} {\inval(i)} } +
               \bEx { \!\sum_{i=s +1}^t\! {\inval(i)} } +
               \bEx {\sum_{i=t}^m {\inval(i)} } \\
            &\leq \O(s) + \O(n^\frac{3}{2} \log^\frac{1}{2}{n}) + \O(n\log{n})
        = \O(n^\frac{3}{2} \log^\frac{1}{2}{n}).
    \end{align*}
    The second bound $\Ex { \sum_{i=1}^m {\inval(i)} } \leq m$ is obvious by
    definition of $\inval(i)$.
\end{proof}


\section{Empirical observations}
\label{sec:exp}

In addition to the achieved average-case bounds,
we also examined AHRSZ and PK experimentally using
the implementation of David J. Pearce~\cite{PK:JEA}
available from www.mcs.vuw.ac.nz/$\tilde\ $djp/dts.html.
For varying number of vertices $n=100,200,\ldots,10000$,
we generated random edge insertion sequences (REIS) leading
to complete DAGs and averaged the performance parameter $C(n)$
over 250 runs.  The chosen $C(n)$ upper bounds the respective runtimes.

\begin{figure}[!h]
    \centering
    \subfloat[$C(n) \ \big/ \ (n^2 \log{n})$]{
        \includegraphics[bb=82pt 328pt 522pt 790pt,angle=-90,width=.47\textwidth,clip]{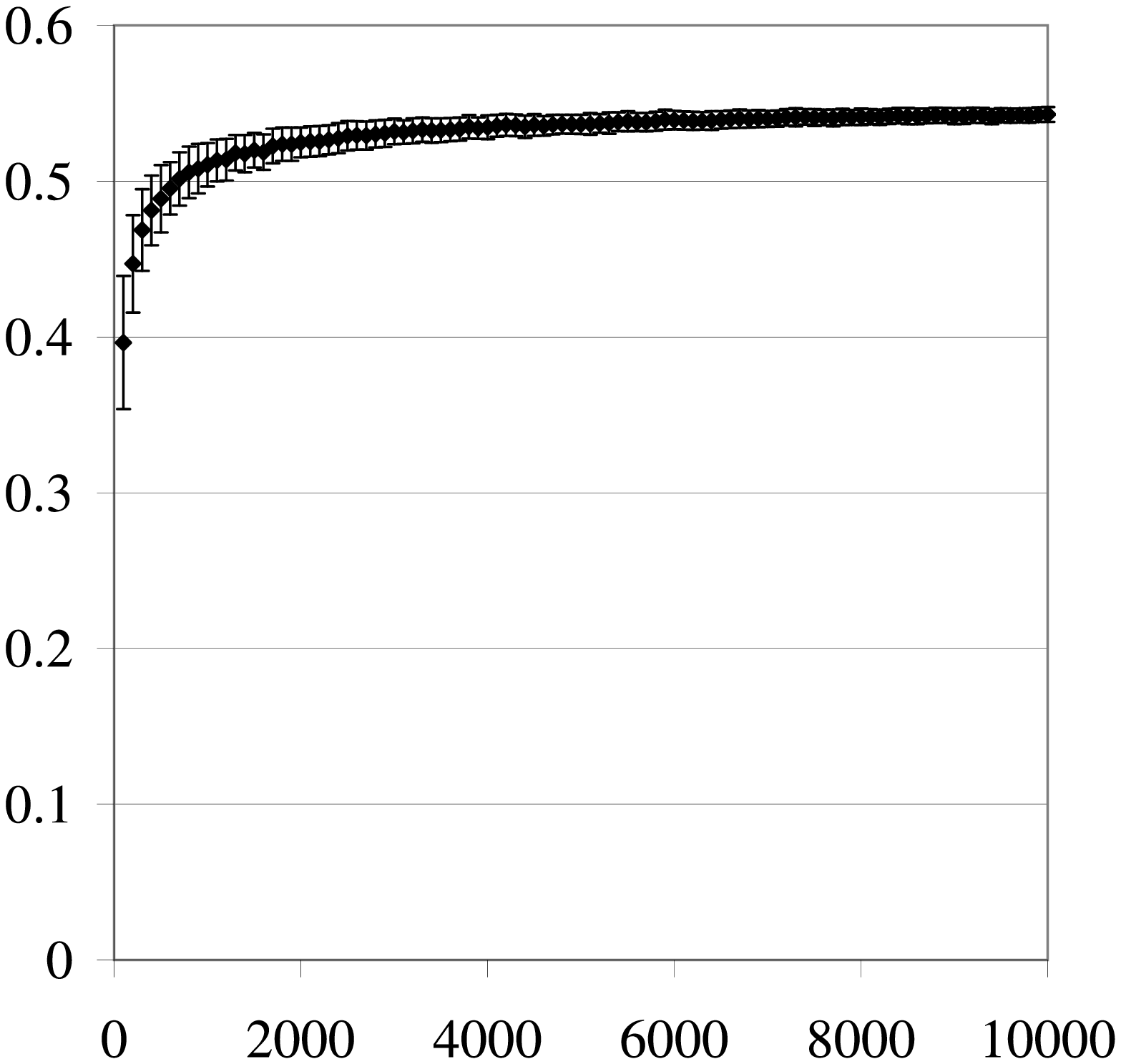}
        \label{subfig:log}
    }
    \subfloat[$C(n) \ \big/ \ (n^2 \log^2{n})$]{
        \includegraphics[bb=82pt 318pt 522pt 780pt,angle=-90,width=.47\textwidth,clip]{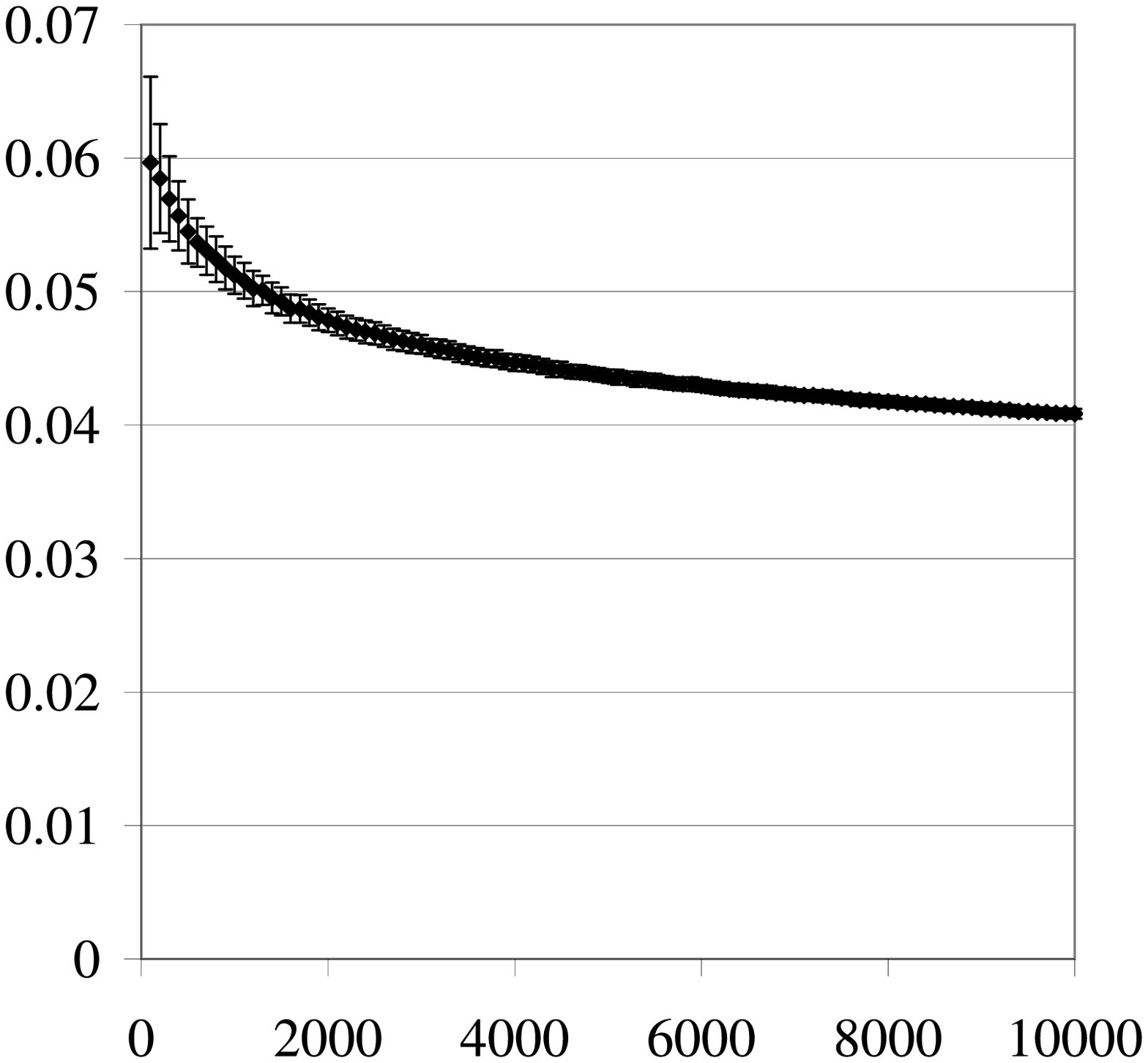}
        \label{subfig:log2}
    }
    \caption{Experimental results of AHRSZ
            for the insertion of the edges of a complete DAG in a random order.
            The horizontal axes describe the number of vertices $n$.
            The vertical axes show the measured 
            empirical insertion costs
            $C(n):=\sum_i \kkkk\, \log\kkkk$
            relative to \protect\subref{subfig:log} $n^2 \log{n}$
            and \protect\subref{subfig:log2} $n^2 \log^2{n}$, respectively.
            The error bars specify the sample standard deviation.
    }
    \label{fig:exp}
\end{figure}

The performance parameter taken for AHRSZ is $C(n):=\sum_i \kkkk\,\log({\kkkk})$.
We know $\bEx{C(n)}=\O(n^2 \log^3{n})$ from \secref{AHRSZ}
and know that the overall runtime is $\Omega(n^2)$ since the algorithm has to inspect all the edges being inserted. 
In our experimental setting, we discovered 
that $C(n)/ (n^2\log^2{n})$ is apparently 
a decreasing function
and that $C(n)/ (n^2 \log{n})$ is an increasing function.
This empirical evidence suggests that $C(n)$ is possibly between $\Omega(n^2 \log{n})$ and $\O(n^2 \log^2{n})$.
\figref{exp} shows our experimental results for AHRSZ.

We consider $C(n):=\sum_i (\dddi +\ddi \log{\ddi})$ as a performance parameter for PK
and observe that $C(n) / n^2$ is decreasing while
$C(n) / (n^2 \log^{-1}{n})$ is increasing.  This indicates that 
$C(n)=o(n^2)$, which implies an actual runtime of $\Theta(n^2)$ for PK on REIS  
since all $\Omega(n^2)$ edges have to be inspected.
\citet{PK:JEA} showed empirically that PK outperforms AHRSZ on sparse DAGs.
Our experiments extend this to dense DAGs.

Complementing \secref{inval}, we also examined empirically the
number of invalidating edges for AHRSZ.  The same experimental set-up as above
suggests a quasilinear growth of $\sum_{i=1}^m {\inval(i)}$ between    
$\Omega(n \log{n})$ and $\O(n \log^2{n})$.  Note that the observed empirical bound
for AHRSZ is significantly lower
than the general bound $\O(n^{\frac{3}{2}} \log^{\frac{1}{2}}{n})$
of \thmref{inval} which holds for all algorithms.


\section{Discussion}

On random edge insertion sequences (REIS) leading to a complete DAG, we 
have shown an expected runtime of $\O(n^2 \log^2{n})$ for PK
and $\O(n^2 \log^3{n})$ for AHRSZ and KB while the trivial lower bound
is $\Omega(n^2)$.
Extending the average case analysis for the case where we only insert $m$ edges with $m \ll n^2$ still remains open.
On the other hand, the only non-trivial lower bound for this problem is by \citet{Low-Bound}, who have shown that an adversary can force 
any algorithm which maintains explicit labels to require $\Omega(n \log{n})$ time complexity for 
inserting $n-1$ edges.
There is still a large gap between the lower bound of
$\Omega(\max\{n \log{n},m\})$, the best average-case bound of $\O(n^2 \log^2{n})$
and the worst-case bound of $\O(\min\{m^{1.5} + n^2 \log{n}, m^{1.5} \log{n}, n^{2.75}\})$.
Bridging this gap remains an open problem.

\section*{Acknowledgements}

The authors are grateful to Telikepalli Kavitha, Irit Katriel, and Ulrich Meyer for various helpful discussions.

\end{document}